\documentclass[a4paper,fleqn]{cas-dc}

\usepackage[numbers]{natbib}

\def\tsc#1{\csdef{#1}{\textsc{\lowercase{#1}}\xspace}}
\tsc{SAXS}
\tsc{DOC}

\begin{document} 

\def\floatpagepagefraction{1}
\def\textpagefraction{.001}
\shorttitle{Multiscale X-ray computed tomography of standard optical fibers}
\shortauthors{M.C. Crocco et~al.}

\title [mode = title]{Multiscale X-ray computed tomography of standard optical fibers}                      



\author[1,2]{M.C. Crocco}[type=editor,
                        auid=000,bioid=1,
                        orcid=0000-0002-9923-2754
                        ]
                        \cormark[1]
                        \ead{mariacaterina.crocco@unical.it}
\credit{Conceptualization, Data curation, Formal analysis, Writing - Original draft preparation}

\author[3,4]{F. Cognigni}[style=editor]
\credit{Data curation, Formal analysis, Writing - Original draft preparation}

\author[3]{A. Sanna}[style=editor]
\credit{Data curation, Writing – review \& editing}

\author[2]{R. Filosa}[style=editor]
\credit{Data curation, Writing – review \& editing}

\author[2]{S. Siprova}[style=editor]
\credit{Data curation, Writing – review \& editing}

\author[1,2]{R.C. Barberi}[style=editor]
\credit{Writing – review \& editing}

\author[1,2]{R.G. Agostino}[style=editor]
\credit{Writing – review \& editing}

\author[5]{S. Wabnitz}[style=editor]
\credit{Writing – review \& editing}

\author[5]{A. D'Alessandro}[style=editor]
\credit{Writing – review \& editing}

\author[6]{S. Lebrun}[style=editor]
\credit{Writing – review \& editing}

\author[3]{M. Rossi}
\credit{Validation, Writing – review \& editing}

\author[1,2]{V. Formoso}[style=editor]
\credit{Validation, Writing – review \& editing}

\author[2]{R. Termine}[style=editor]
\credit{Validation, Writing – review \& editing}

\author[1,2,7,8]{A. Bravin}
\credit{Data curation, Validation, Writing – review \& editing}

\author[1,2]{M. Ferraro}[type=editor,
                        auid=000,bioid=1,
                        orcid=0000-0002-6014-2890
                        ]


\credit{Conceptualization, Writing - Original draft preparation, Supervision}

\affiliation[1]{organization={Department of Physics and STAR Research Infrastructure, University of Calabria},
                addressline={Via Tito Flavio}, 
                city={Rende},
                postcode={87036}, 
                state={(CS)},
                country={Italy}}
\affiliation[2]{organization={CNR-NANOTEC, SS di Rende},
                addressline={Via Pietro Bucci}, 
                city={Rende},
                postcode={87036}, 
                state={(CS)},
                country={Italy}}
\affiliation[3]{organization={Department of Basic and Applied Sciences for Engineering (SBAI), University of Rome La Sapienza},
                addressline={Via Antonio Scarpa 14}, 
                city={Rome},
                postcode={00161}, 
                country={Italy}}

\affiliation[4]{organization={Carl Zeiss S.p.A, Research Microscopy Solutions},
                addressline={Via Varesina, 162}, 
                city={Milan},
                postcode={20156}, 
                country={Italy}}

\affiliation[5]{organization={Department of Information Engineering, Electronics, and Telecommunications, Sapienza University of Rome},
                addressline={Via Eudossiana 18}, 
                city={Rome},
                postcode={00184}, 
                country={Italy}}

\affiliation[6]{organization={Université Paris-Saclay, Institut d’Optique Graduate School, CNRS},
                addressline={Laboratoire Charles Fabry}, 
                city={Palaiseau},
                postcode={91127}, 
                country={France}}
                
\affiliation[7]{organization={Department of Physics, “G. Occhialini”, University Milano Bicocca},
                addressline={Piazza della Scienza 3}, 
                city={Milan},
                postcode={20126}, 
                country={Italy}}

\affiliation[8]{organization={European Synchrotron Radiation Facility},
                addressline={ 71 Avenue des Martyrs}, 
                city={Grenoble},
                postcode={38000}, 
                country={France}}

\cortext[cor1]{Corresponding author}


\begin{abstract}
Optical fiber technologies enable high-speed communication, medical imaging, and advanced sensing. Among the techniques for the characterization of optical fibers, X-ray computed tomography has recently emerged as a versatile non-destructive tool for mapping their refractive index variations in 3D. In this study, we present a multiscale characterization of standard optical fibers. We carry out an intercomparison of three tomography setups: classical computed microtomography, X-ray microscopy, and nanotomography. In each method, our analysis highlights the trade-offs between resolution, field of view, and segmentation efficiency. Additionally, we integrate deep learning segmentation thresholding to improve the image analysis process. Thanks to its large field of view \textcolor{black}{(10 x 10 mm\textsuperscript{2})}, microtomography with classical sources is ideal for the analysis of relatively long fiber spans, where a low spatial resolution is acceptable. The other way around, nanotomography has the highest spatial resolution \textcolor{black}{(50 - 150 nm)}, but it is limited to very small fiber samples, e.g., fiber tapers and nanofibers\textcolor{black}{, which have diameters of the order of a few microns}. Finally, X-ray microscopy provides a good compromise between the sample size \textcolor{black}{(of the order of 1 mm)} fitting the device's field of view and the spatial resolution needed for properly imaging the inner features of the fiber \textcolor{black}{(about 1 $\mu$m)}. Specifically, thanks to its practicality in terms of costs and cumbersomeness, we foresee that the latter will provide the most suitable choice for the quality control of fiber drawing in real-time, e.g., using the "One-Minute Tomographies with Fast Acquisition Scanning Technology" developed by Zeiss. In this regard, the combination of X-ray computed tomography and artificial intelligence-driven enhancements is poised to revolutionize fiber characterization, by enabling precise monitoring and adaptive control in fiber manufacturing \textcolor{black}{(such as fiber size and non-circularity)}. 

\end{abstract}


\maketitle

\section{Introduction}


Optical fiber-based technologies have widespread applications in modern communications, medicine, and industry. From the high-speed internet to life-saving medical imaging and advanced sensing systems, they enable the fast and reliable transmission of information that keeps our world connected.
In short, a standard optical fiber consists of a core and a surrounding cladding, both made of glass. The core has a higher refractive index than the cladding, allowing light to be guided through the mechanism of total internal reflection. In silica fibers, such a refractive index contrast is achieved either by doping the core with heavier elements, such as germanium, or by doping the cladding with lighter elements, such as fluorine.

Beyond standard silica fibers, novel materials such as soft glass and polymers are being explored to push the boundaries of fiber capabilities and performance. For instance, soft glass fibers offer transmission in the mid-infrared spectral range, where standard fibers are limited owing to silica high absorption, while polymer optical fibers provide greater flexibility and cost-effectiveness, making them ideal for short-distance communications and wearable internet-of-things devices \cite{ferreira2025roadmap}. In addition, optical fiber tapers and nanofibers have been developed for sensing technologies that leverage enhanced light-matter interactions through strong evanescent fields and subwavelength confinement \cite{godet2017brillouin,lebrun2024design,beugnot2014brillouin}.

Interestingly, although optical fiber technologies date back to the 1960s, the characterization of optical fibers, e.g., the profiling of the core and the cladding refractive index, still presents some challenges \cite{stewart1982optical}. Most of the commonly used techniques provide information about the refractive index of the fiber facet, disregarding possible variations that may occur during fiber drawing. These are the so-called longitudinal techniques, since the probe is parallel to the fiber axis \cite{zhong1994characterization,young1981optical,ikeda1975refractive, weng2015exploiting,presby1976refractive}. On the other hand, transverse techniques, where the probe is perpendicular to the fiber axis, may provide information about the refractive index profile along the fiber axis. This is an invaluable feature when one wants to control the stability of fiber drawing \cite{sochacka1994optical,barty1998quantitative,yablon2009multi,bachim2005microinterferometric,gorski2007tomographic,fan2020reconstructing}.

Among the many transverse techniques, those based on X-rays have recently gained significant interest. For instance, when applied to polymer optical fibers, small-angle X-ray scattering provides insights into polymer chain orientation, phase separation, and structural inhomogeneities, which influence the optical and mechanical properties of the fibers \cite{beckers2017novel,lee2022small,perret2022effects,ferraro2025small}. Whereas, when applied to silica optical fibers, X-ray-based techniques were used to study the density and concentration fluctuations of dopants \cite{le2008density}. However, the most versatile X-ray-based technique for characterizing optical fibers is, with no doubt, X-ray computed tomography (XCT). In fact, this technique permits to non-destructively retrieve a 3D map of the refractive index profile, regardless of the material and the presence of fiber bending or coating with plastic materials \cite{sandoghchi2014x,levine2019multi,levine2021x, ferraro2022x, crocco2024soft}.

In the case of pure absorption contrast tomography, one may assume that the XCT intensity at X-ray frequencies ($I_\mu$) takes the shape of the fiber core index profile $n$ at optical frequencies. This is because the material density, which is proportional to $I_\mu$, is determined by the doping concentration, which, in turn, is proportional to $n$. In short, a flat profile of $I_\mu$ in the core corresponds to a step-like profile of the refractive index. Whereas a smooth variation of $I_\mu$ when passing from the core to the cladding indicates the presence of a graded refractive index. This concept has been tested in many types of optical fibers \cite{ferraro2022x}, and turns out to be particularly effective with fibers doped with heavy elements, where the absorption contrast is higher \cite{crocco2024soft}.

In this work, we present a multiscale approach to the XCT of standard optical fibers. Specifically, we scanned and analyzed a given set of optical fiber samples with three different techniques and using classical sources: computed microtomography ($\mu$CT), X-ray microscopy (XRM), and computed nanotomography (nCT). Each of these techniques has its own advantages and disadvantages in terms of resolution, acquisition time, size of the field of view, constraints on the sample size, costs, phase-contrast, and customizability.
This work aims to provide a fair inter-comparison among these techniques when applied to optical fibers. To this goal, we dedicate Sec. \ref{sec:meth} to presenting the used materials and acquisition methods. In Sec. \ref{sec:ris} we discuss the results of the analysis of the XCT images. We report the 3D rendering and the $I_\mu$ maps, focusing on the effects given by the different resolutions of the three techniques. We also introduce artificial intelligence (AI)-assisted segmentation, which efficiently supports the segmentation of XCT images in the presence of artifacts. Finally, we compare the results obtained in terms of the fibers' geometrical parameters, before concluding in Sec. \ref{sec:con}.

\section{Materials and Methods}
\label{sec:meth}

For our investigation, we considered commercial fibers, chosen among the most commonly used. Besides the wider range of users that might be interested in these results, this choice is driven by the fact that the industrial processes involved in the drawing of several kilometers of fiber may last for a relatively long time, thus being more likely affected by environmental perturbations. We used three standard Thorlabs fiber spans: a step-index (henceforth STEP) multimode fiber, product code FG050LGA, a graded-index (GRIN) fiber, product code GIF50E, and a single-mode (SMF) fiber span, product code SMF-28-J9. The nominal values of the core diameter ($D_{core}$), cladding diameter ($D_{clad}$), core non-circularity (\textit{NC}${}_{core}$), and cladding non-circularity (\textit{NC}${}_{clad}$) are reported in Table \ref{tab:samples}. According to the manufacturer, the STEP fiber has a pure silica core and fluorine-doped cladding. Whereas the SMF and the GRIN fibers have a pure silica cladding and a Germanium-doped core. The manufacturer provides neither the values of the doping nor those of the refractive index. However, it is indicated that the GRIN fiber's core has a bell-shaped refractive index profile.

\begin{table}[!hb]
    \centering
    \begin{tabular}{ccccc}
\toprule
       Fiber & $D_{core}$ ($\mu$m) & $D_{clad}$ ($\mu$m) & \textit{NC}${}_{core}$ & \textit{NC}${}_{clad}$ \\
\midrule
    STEP & 50.0 $\pm$ 2.5 & 125.0 $\pm$ 0.5 & - & -\\
       GRIN & 50.0 $\pm$ 1.0 & 125.0 $\pm$ 1.0 & $\leq 5\%$ & $\leq 1\%$\\
SMF  & - & 125.0 $\pm$ 0.7 & - & -\\

\bottomrule
    \end{tabular}
    \caption{Geometric parameters of the samples as provided by the manufacturer.}
    \label{tab:samples}
\end{table}

Finally, for nCT, we used a taper of an SMF from Alcatel, product code P25Z0175E (2003). The Alcatel and Thorlabs SMF have the same nominal parameters, both in terms of their geometrical and optical characteristics. The taper was made in-house, at the Paris-Saclay University, France, by pulling the two ends of the fiber, which were attached to two computer-controlled translation stages. These elongate the fiber to create the taper under the action of a heater. In this way, we could provide the taper with its canonical truncated cone shape \cite{birks1992shape,baker2011generalized}.

The experimental parameters of the device we used are reported in Table \ref{tab:parameters}. 
\begin{table*}[!b]
    \centering
    \begin{tabular}{cccc}
\toprule
         & $\mu$CT & XRM & nCT \\
\midrule
       Energy (keV) & 60 (max) & 60 (max) & 5.4 \\
       Tube voltage (kV) & 60 & 60 & 35 \\
       Tube current (mA) & 0.166 & 0.108 & 25 \\
       Acquisition time (s) & 3.50 & 15 & 60 \\
       Magnification  & 11.5 & 14.1 & 203.1 \\
       Pixel size at sample position ($\mu$m) & 4.340 & 0.950 & 0.064 \\
       Filter & Al 40 $\mu$m & LE1 & No filter \\
       Nos. of projection & 3601 & 3201 & 2601 \\
       Angular step (°) & 0.100 & 0.056 & 0.069\\
       Image size ($pixel$ x $pixel$) & 2240x2368 & 1024x1024 & 1024x1024 \\
       Detector pixel size ($\mu$m) & 50.0 & 13.5 & 13\\
       Field of view ($\mu m^2$) & 9721x10227 & 973x973 & 65x65 \\
\bottomrule
    \end{tabular}
    \caption{Experimental parameters}
    \label{tab:parameters}
\end{table*}

X-ray microtomography measurements were performed at the $\mu$tomo1 laboratory of STAR Research Infrastructure in Rende, Italy. The experimental apparatus consists of a microfocus source (Hamamatsu L12161-07), a flat panel detector (Hamamatsu C7942SK-05), and a handling system with 5 degrees of freedom (one rotator, two goniometer cradles, and two handling motors orthogonal to the beam direction). The system is designed to change the source-object distance (SOD) and source-detector distance (SDD) by keeping the object position fixed and moving the source and the detector. \textcolor{black}{A sketch of the apparatus can be found in Ref. \cite{ferraro2022x}.}
The acquisition parameters were carefully calibrated to obtain high-quality images of the samples under investigation. 
To improve the signal-to-noise ratio (SNR), for each angular step, 3 projections were acquired and averaged. The acquired projections were normalized by using flat and dark images from the Fiji ImageJ's ANKAphase v. 2.1 plugin, and then reconstructed with the Nrecon v. 1.7.4.6 software using the Feldkamp-Davis-Kress (FDK) algorithm, after correcting beam hardening and ring artifacts \cite{schindelin2012fiji, weitkamp2011ankaphase, nrecon, feldkamp1997practical}. We underline that the $\mu$CT analysis of the STEP and GRIN fibers was already reported in one of our former works \cite{ferraro2022x}.

As far as XRM is concerned, the datasets analyzed in this paper were acquired using a ZEISS Xradia Versa 610 X-ray microscope at the Research Center on Nanotechnology Applied to Engineering of Sapienza (CNIS), located at La Sapienza, University of Rome \cite{cognigni2023exploring,cognigni2023multimodal}. \textcolor{black}{A sketch of the device can be found in \cite{versa}.}
This instrument features a high photon flux source, enabling a stronger X-ray signal, faster scans, and an improved SNR. Its two-stage magnification architecture, which combines geometric and optical magnification, achieves sub-micrometer resolution.
To perform multiscale XRM experiments, the Scout-and-Zoom procedure was employed. This approach allows the selection of volumes of interest (VOIs) within a low-resolution scan to be subsequently scanned at higher resolution. 
The XRM dataset was automatically reconstructed using the ZEISS Scout-and-Scan Control System Reconstructor (V.16.1.14271.44713) with the traditional FDK cone-beam reconstruction algorithm.

Finally, nCT measurements were conducted by using the Zeiss Xradia 810 Ultra system \cite{ultra}, available at the Institute of Nanotechnology labs of CNR, hosted within the STAR Research Infrastructure in Rende, Italy. The source's anode exploits the K$_\alpha$ emission of Chromium, providing quite low energy photons. 
Despite its relatively low photon flux, this device provides high brightness, thanks to the focusing of the X-ray beam onto the sample via a high-efficiency capillary condenser. Moreover, Fresnel zone plate objectives image the transmitted X-rays onto the detector, ensuring a very high spatial resolution.
The device may also allow for obtaining Zernike phase-contrast images by inserting a phase ring into the beam path, thus enhancing the visibility of features in low-absorbing samples such as optical fibers. However, we chose to work with a pure absorption-contrast setting, since the occurrence of phase contrast would invalidate the hypothesis that the XCT intensity provides a map of the fiber refractive index.

\begin{figure}
    \centering
    \includegraphics[width=0.99\linewidth]{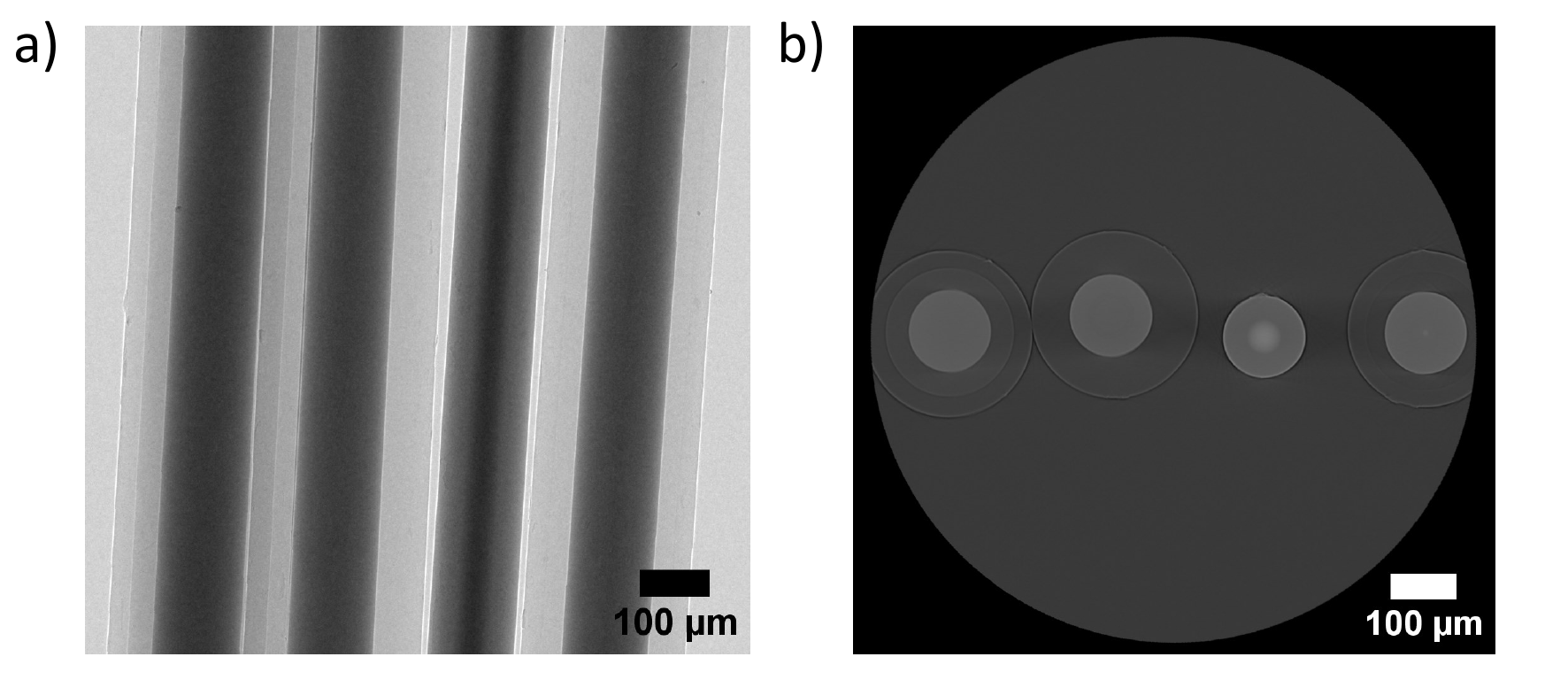}
    \caption{\textcolor{black}{Arrangement of optical fibers in the XRM system. a) Radiographic image of the optical fibers. b) Tomographic section.}}
    \label{fig:position}
\end{figure}

Before presenting the results, it is important to mention that, in the case of $\mu$CT and nCT, each fiber was scanned individually. To the contrary, in the case of XRM, all samples were put together in a single round of measurement\textcolor{black}{, as shown in Fig. \ref{fig:position}}. \textcolor{black}{The reason for scanning multiple samples in a single shot was to save time. As a matter of fact, the acquisition time needed to perform a single tomography, regardless of the number of samples, was as long as half a day (that is, 15 s times 3201 projections, cfr. Table \ref{tab:parameters}).}

\section{Results}
\label{sec:ris}

\begin{figure*}[!hb]
\centering
\includegraphics[width=15.5cm]{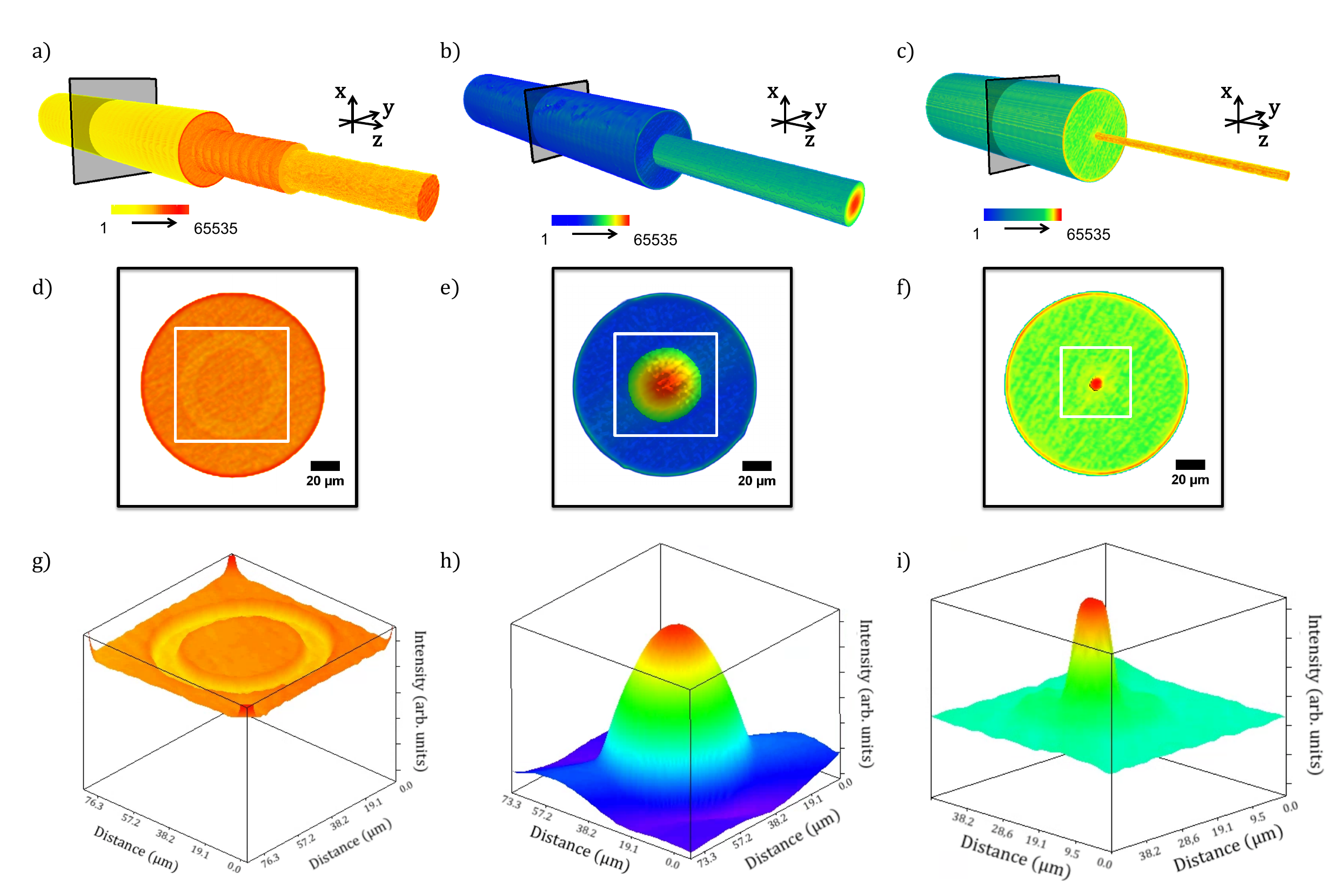}			
\caption{XRM tomography results. (a-c) False color 3D rendering of the STEP (a), GRIN (b), and SMF (c) samples, respectively. The color bars represent the grayscale intensity that grows according to the arrow's direction. \textcolor{black}{The grayscale is in arbitrary units and ranges between 0 and 65535, that is $2^{16}-1$, as the images are in 16 bits. The value 0 corresponds to the white background.} (d-f) Color plot of the single slices indicated by the dark squares in (a-c), respectively. (g-i) Surface plot of the regions of interest in (d-f) indicated by white squares.}	
\label{fig:3Dplot}%
\end{figure*}

After their reconstruction, the analysis of the tomographic images was carried out by using Fiji ImageJ, Arivis Cloud, and Dragonfly. Fiji ImageJ was used to isolate each optical fiber (in the case of XRM tomography) and to vertically align the fiber axis. Arivis Cloud was used, instead, to train and apply the deep learning semantic segmentation model. Dragonfly 3D World 2024.1 was used for the visualization of 3D renderings and 2D tomography slices. Finally, we used IgorPro 9 for plotting the results of our analysis.

\subsection{3D rendering and profile}

The 3D rendering of the XRM tomographies of the STEP, the GRIN, and the SMF samples is shown in Fig. \ref{fig:3Dplot}a-c, respectively. Here, we dub $z$ the coordinate in the direction parallel to the fiber axis; whereas $xy$ are the coordinates in the plane orthogonal to the fiber axis. As can be seen, there is a marked difference between the three samples: this is emphasized in the geometric segmentation shown in Fig. \ref{fig:3Dplot}a-c. Here, we virtually separate the core from the cladding \textcolor{black}{by the following process: the tomographic images were segmented to assign a gray value to a given element (e.g., core and cladding); for each element a binarization mask was obtained by assigning the value 1 to the region of interest, and value 0 to the remaining part of the image; finally, for each given element, the original images were multiplied by the binarization mask. In this way, we preserved the original gray levels.} 

In Fig. \ref{fig:3Dplot}d-f, we show a 2D image of the whole fiber section, i.e., a single $xy$ slice, as indicated by the dark squares in Fig. \ref{fig:3Dplot}a-c. The color plots in Fig. \ref{fig:3Dplot}d-f have the same spatial scale, to allow for a fair naked-eye comparison of the core and cladding sizes of all samples. Moreover, since most of the features of our interest are located in the fiber core, we defined a region of interest that includes the core and its vicinity, as indicated by the white squares in Fig. \ref{fig:3Dplot}d-f. The resulting intensity map is shown in Fig. \ref{fig:3Dplot}g-i. 

The STEP fiber shows a virtually flat XCT intensity in its core (see the orange facet in Fig. \ref{fig:3Dplot}a and the inner circle in Fig. \ref{fig:3Dplot}d,g). On the other hand, the GRIN fiber has a smooth variation of the intensity when moving from the fiber axis to the core edges (cfr. Fig. \ref{fig:3Dplot}b,e,h). As far as the SMF is concerned, it turns out to have a graded intensity map, similar to that of the GRIN fiber (cfr. Fig. \ref{fig:3Dplot}i and \ref{fig:3Dplot}h). This property is of extreme interest whenever operating the SMF fiber at wavelengths below its cutoff value, e.g., when dealing with supercontinuum generation applications. 
Finally, we found an intermediate layer, the so-called fiber trench, in both the STEP and SMF samples. It is interesting to point out that both the graded nature of the SMF index profile and the presence of the trench were not mentioned in the manufacturer's specs sheet. 

\begin{figure}
    \centering
    \includegraphics[width=0.85\linewidth]{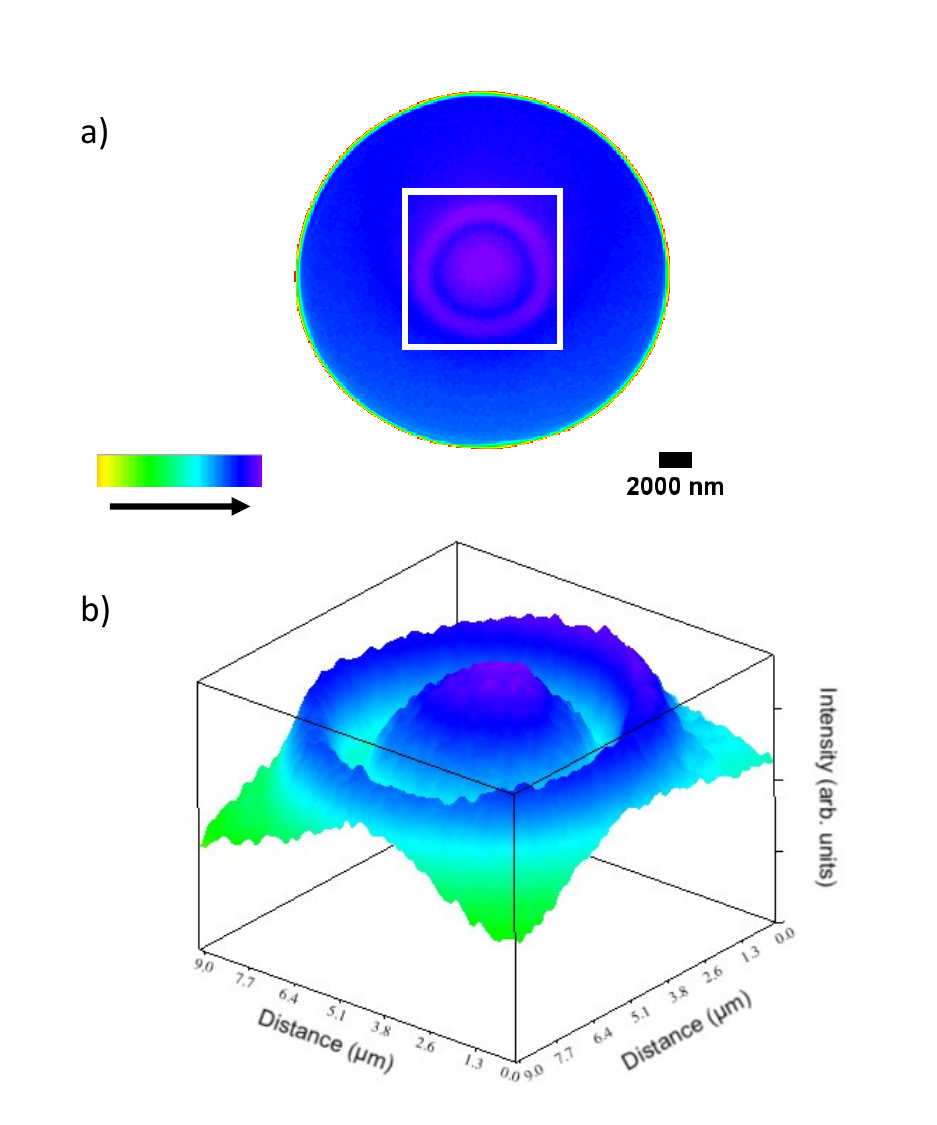}
    \caption{nCT analysis of the SMF taper sample. (a) Color plot of a single $xy$ slices. The color bars represent the grayscale intensity that grows according to the arrow's direction. (b) Surface plot of the region of interest in (a) indicated by the white square.}
    \label{fig:nCT}
\end{figure}

The results of nCT are shown in Fig. \ref{fig:nCT}. In analogy to Fig. \ref{fig:3Dplot}, we show a single $xy$ slice and a region of interest including the fiber core in Fig.\ref{fig:nCT}a and b, respectively. As can be seen, the core of the taper sample has a parabolic-like profile. In addition, we could clearly spot the presence of a trench.


\subsection{Spatial resolution and field of view}
\label{sec:res}

One of the most important parameters when dealing with tomography in general, and with the tomography of optical fibers in particular, is the spatial resolution. \textcolor{black}{This depends on several factors, such as the modulation transfer function \cite{villarraga2020amplitude}. Still,} at first approximation, the spatial resolution \textcolor{black}{could be considered} proportional to the equivalent pixel size. 

\begin{figure}
\centering
\includegraphics[width=8.0cm]{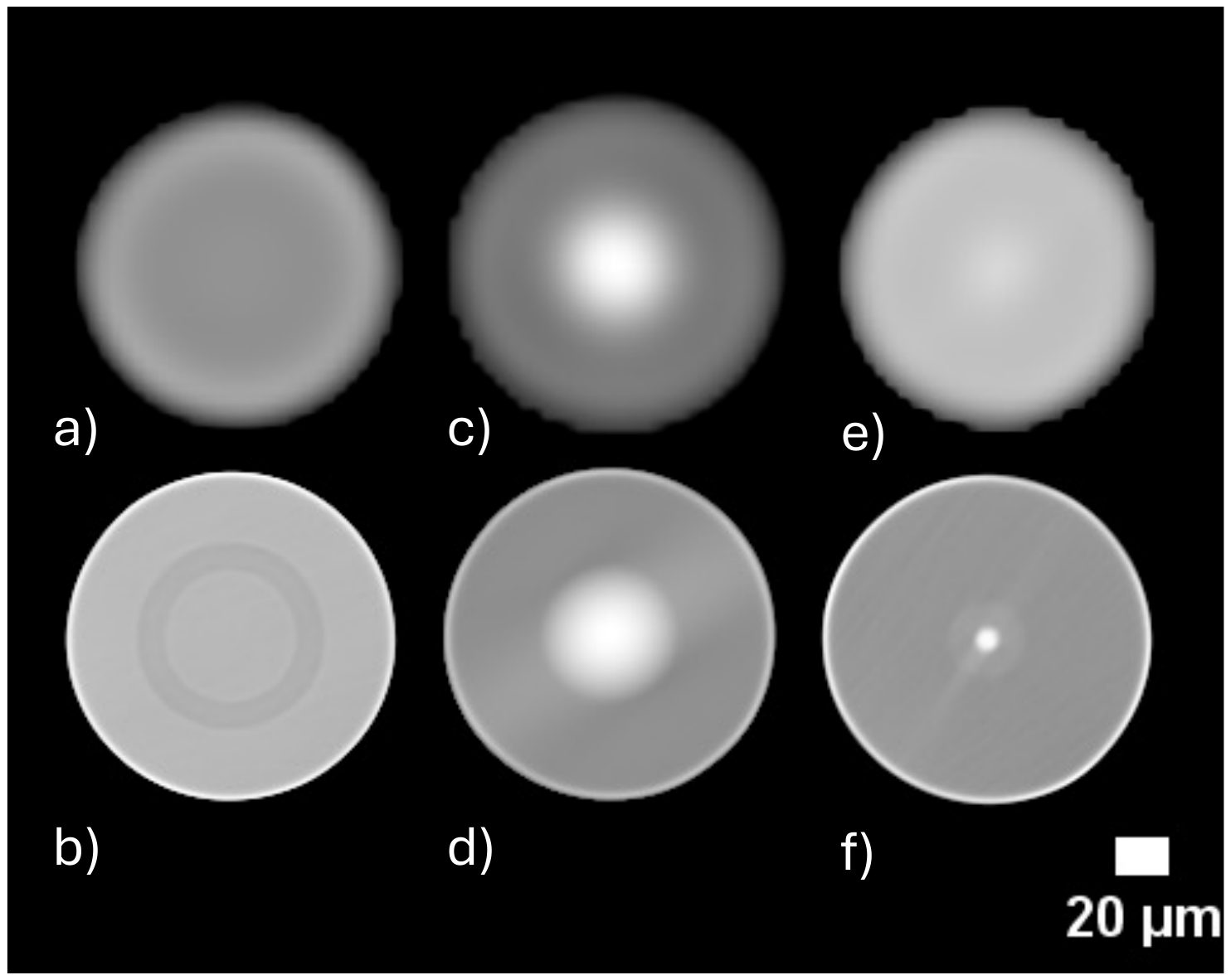}			\caption{Comparison between the $xy$ slices of the STEP (a,b), the GRIN (c,d), and the SMF (e,f) samples obtained via $\mu$CT and XRM, respectively. 
}
\label{fig:confronto}%
\end{figure}

The values of the detector pixel size reported in Table \ref{tab:parameters} are adequate for all of the XCT techniques whenever large core fibers, such as the STEP and GRIN multimode fibers used in this work, are involved. However, when it comes to the SMF, whose typical core diameter is below 10 $\mu$m, the $\mu$CT is unable to properly distinguish the core from the cladding. 
A comparison of $\mu$CT and XRM applied to the same set of samples (GRIN, STEP, and SMF) is shown in Fig. \ref{fig:confronto}. As can be seen, in the XRM image one may clearly identify the fiber core as well as the trench. Whereas, in $\mu$CT images the core appears blurred. On the other hand, as far as the cladding diameter is involved, $\mu$CT provides similar results to XRM.

In terms of resolution, nCT is the technique with the highest performances. This can be easily understood by visually compare Fig. \ref{fig:3Dplot}f,i with Fig. \ref{fig:nCT}a,b. Indeed, one may appreciate much finer details in the core of the fiber samples when using nCT rather than XRM. Nonetheless, nCT has a major drawback when applied to standard optical fibers, i.e., its limited field of view. For instance, the field of view of our nCT device, which is about 65 $\mu$m x 65$\mu$m (see Table \ref{tab:parameters}), is too small to image standard optical fibers whose cladding diameter is 125 $\mu$m. Furthermore, the photon energy of our nCT facility is very low, thus making virtually impossible to penetrate thick samples with high atomic number.


\subsection{Deep learning-assisted segmentation}
\label{sec:deep}

\begin{figure}
\centering
\includegraphics[width=8.0cm]{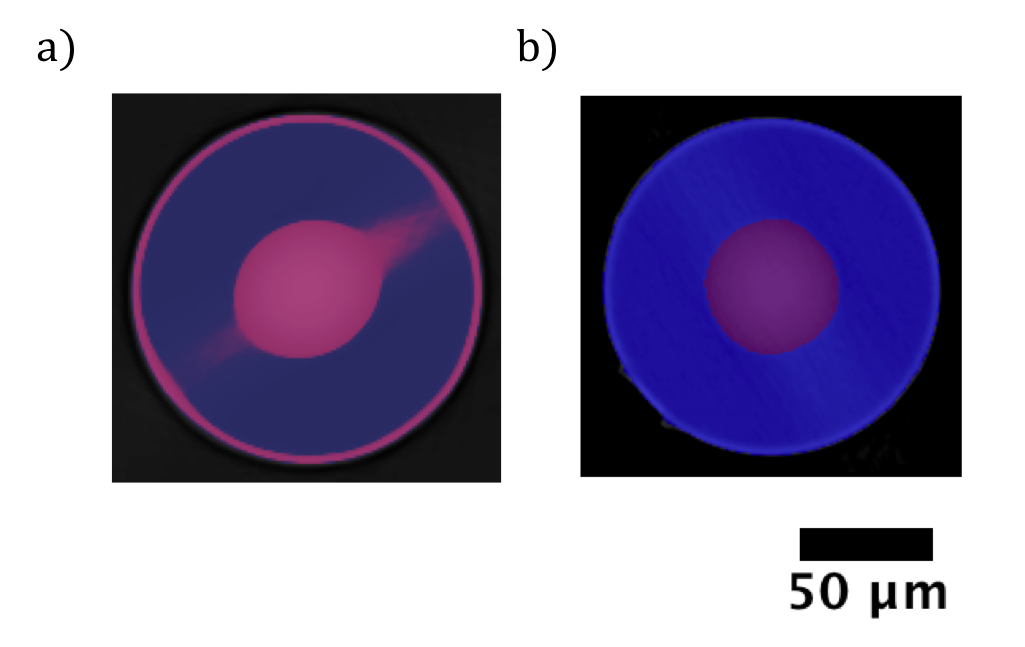}
\caption{Comparison between manual and DL-assisted segmentation of the XRM image of the GRIN fiber. The purple and the blue areas refer to pixels whose associated gray values are above and below the segmentation threshold, respectively.}
\label{fig:thr}%
\end{figure}

The simultaneous measurement of several samples via XRM leads to the formation of artifacts that are detrimental to image analysis. A notable example is the estimation of the core diameter by using histogram-based thresholding. Owing to the presence of the shadow introduced by other samples, the histogram-based thresholding approach used for segmentation returns an artifact, namely a fiber core with an elongated shape. \textcolor{black}{The formation of the shadow artifacts is ascribable to photon starvation effects, which originate from the sample position arrangement that leads to a large difference in absorption between orthogonal projections. The presence of the shadow artifact} is shown in Fig. \ref{fig:thr}a in the case of the GRIN fiber image obtained with XRM. To overcome this issue, we relied on AI-assisted tools, and specifically deep learning (DL) methods. The result of the DL-based segmentation is shown in Fig. \ref{fig:thr}b on the same image as in Fig. \ref{fig:thr}a. As it can be seen, in contrast to the histogram-based thresholding approach in Fig. \ref{fig:thr}a, we obtained a rather circular shape both for the fiber core (purple region) and for the cladding (blue region). This allows a meaningful estimation of the fiber geometrical parameters, which will be discussed in Sec. \ref{sec:geo}. 
We underline that the DL-assisted segmentation was of utmost utility when dealing with the XRM images. As far as $\mu$CT is concerned, since the images are free of artifacts, the use of DL was somehow pleonastic: the manual threshold provided virtually the same outcomes. On the other hand, the DL assistance was not able to improve the quality of the nCT images, i.e., in all cases, the segmentation was significantly hindered by the presence of some artifacts. Notably, we found that the gray levels were not uniform along neither $x$, $y$, nor $z$ (as can be spotted in Fig. \ref{fig:nCT}). This is ascribable to the non-monochromaticity, and to the low energy of the beam combined with the high density of the sample material.

\begin{figure*}[!htb]
\centering
\includegraphics[width=16.0cm]{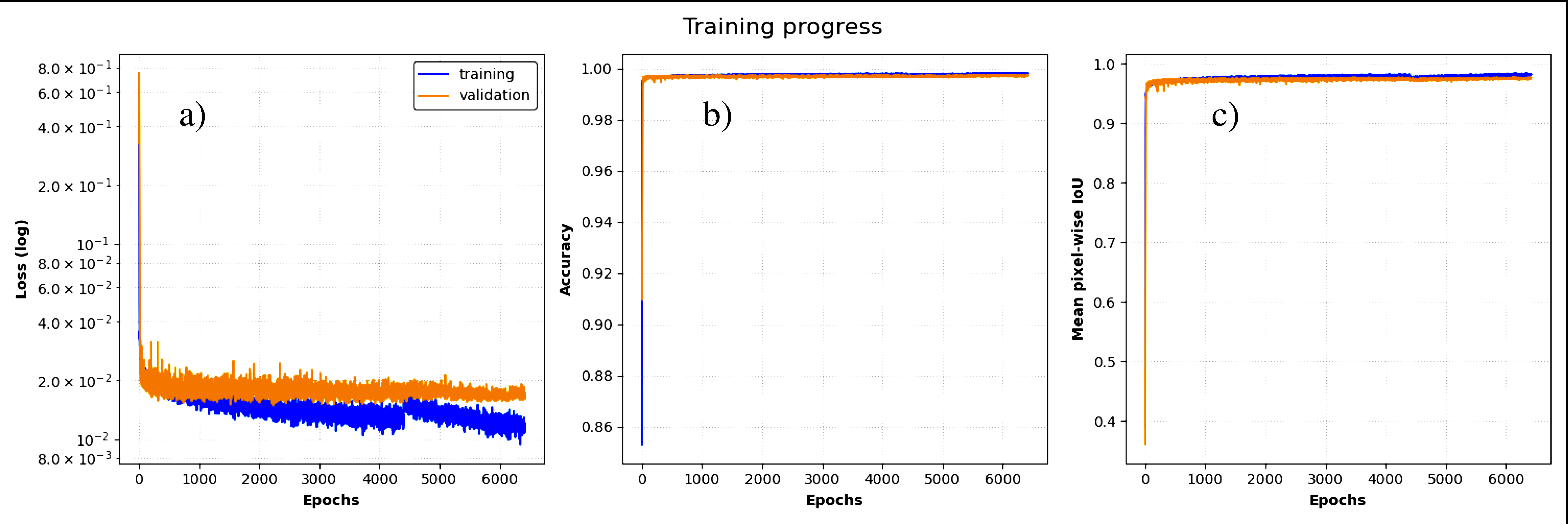}
\caption{Training progresses curves used for DL-assisted segmentation: loss (a), accuracy (b), and mean pixel-wise IoU (c).}
\label{fig:DL}%
\end{figure*}

\begin{figure*}[!hb]
\centering
\includegraphics[width=16.0cm]{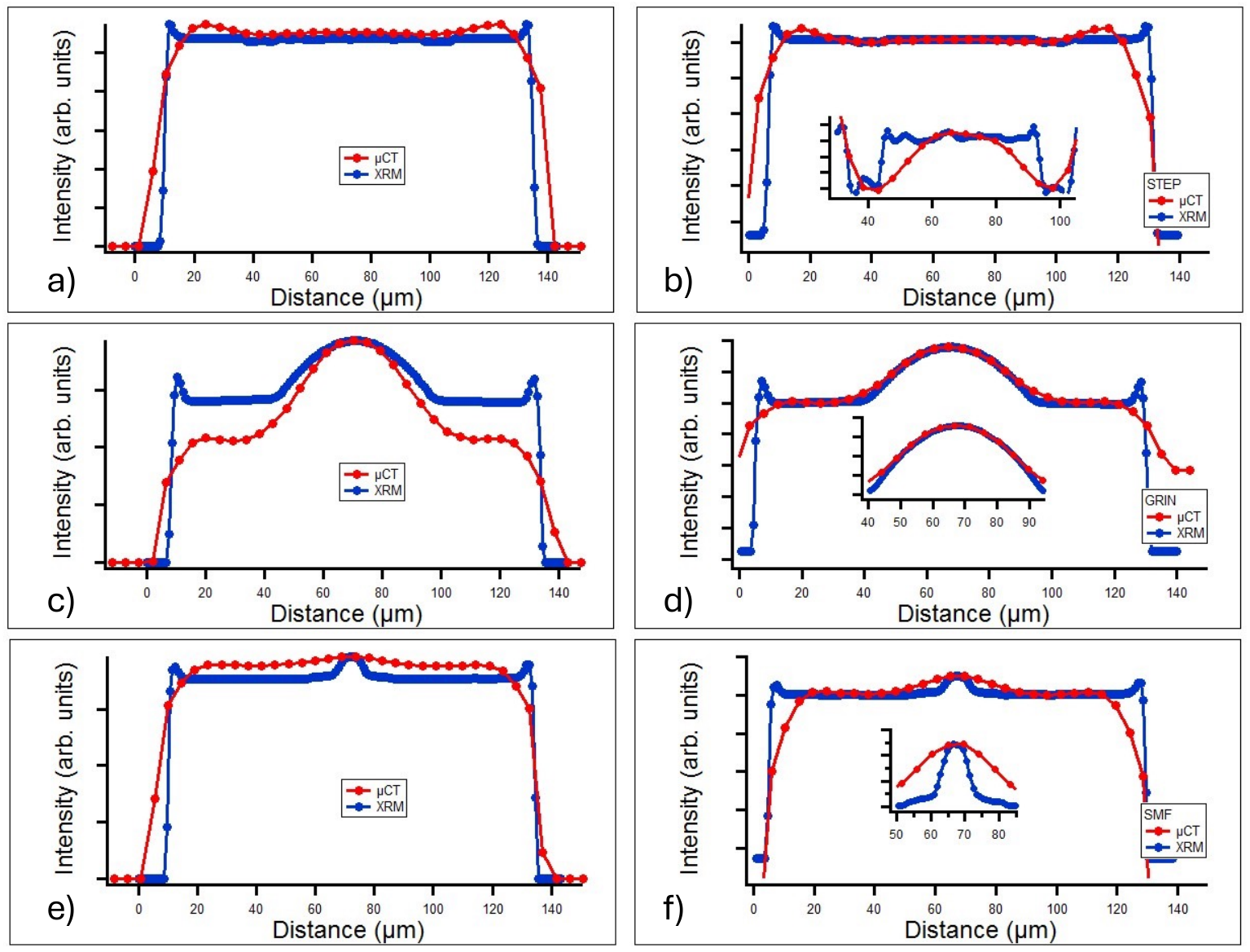}
\caption{XCT intensity profiles for the STEP (a,b), the GRIN (c,d), and the SMF (e,f) samples, respectively. The curves in (a,c,e) were obtained by normalizing the $I_\mu$ profiles to have the same contrast between the core and the air. The curves in (b,d,f) were obtained after cropping and normalizing the $I_\mu$ profiles to have the same contrast between core and cladding.
}	
\label{fig:diam_prof}%
\end{figure*}

The DL-based segmentation models were trained by using cloud computational resources via ZEISS arivis Cloud, which operates on the Microsoft Azure platform \cite{cognigni2024integrating,zeiss}. This solution enables the training of both semantic (pixel-based) and instance (object-based) segmentation models. By annotating a minimal number of objects of interest per class, such as 50 objects, within the required number of images extracted from the entire dataset, it becomes possible to initiate the model training process. In this case, we chose semantic (pixel-based) segmentation to investigate the XRM dataset. Once the model is trained, the dataset can be uploaded to the cloud to perform the desired segmentation. Alternatively, the trained model can be downloaded and integrated into automated local image analysis pipelines, enabling the segmentation process to be executed locally and combined with other image processing and analysis steps.

Input data for training were divided into two datasets, one used for training the model and the other as ground truth for validation. In the plots reported in Figure \ref{fig:DL}, the orange curve represents the validation data, while the blue one indicates the training data. These plots represent loss, accuracy, and mean pixel-wise intersection over union (IoU). Each quantity is plotted as a function of the epoch’s number (an epoch is the time after which the training sees all the training data). Factors, such as the size of the data, determine the total number of epochs. 

The goal of the DL algorithm is to minimize the loss function, which reflects the error between the prediction and the ground truth that was provided. A good model training is witnessed by the loss curve trend being downward for each epoch, and eventually flattening out. This is the case of the loss curve reported in Figure \ref{fig:DL}a. The accuracy is an important metric that the DL algorithm uses for assessing the quality of the model during training. Therefore, the training curve should trend upward, indicating that the model quality is improved at each epoch, as shown in Figure \ref{fig:DL}b. However, the accuracy does not provide information on the quality of segmentation for each of the classes and background in the images. Mean pixel-wise IoU is more recommended for assessing the quality of semantic or instance segmentation. Similarly to the accuracy, IoU is expected to trend upwards as a function of number of epochs, as shown in Figure \ref{fig:DL}c. It is noteworthy that one should consider the overall trend rather than occasional local variations between epochs. Indeed, the curves may look smooth or bumpy, depending on the quality and quantity of the training data.

\subsection{Geometrical parameters}
\label{sec:geo}

We now aim to estimate the geometrical parameters of the fibers, i.e., the diameter and the non-circularity of their core and cladding. Let us first consider the case of the diameters. 
A simple way of determining $D_{core}$ and $D_{clad}$ consists of measuring the distance between the inflection points of the $I_{\mu}$ profile, which are shown in Fig. \ref{fig:diam_prof}. In short, one identifies the inflection points as the minima of the absolute values of the second derivative of $I_\mu$. This method turned out to be quite efficient in Ref. \cite{crocco2024soft}.
It is important to underline that the extraction of the $I_\mu$ profile as in Fig. \ref{fig:diam_prof} was done along a line orthogonal to the streak-like artifact in the case of XRM. Moreover, as it only involves differential quantities, this method was virtually unaffected by the artifacts of nCT.

In Table \ref{tab:diam_prof} we report the average values of $D_{core}$ and $D_{clad}$ along with their standard deviation calculated over 300 slices. Note that such a spatial averaging corresponds to different fiber lengths in each technique, since XRM, $\mu$CT, and nCT experiments were carried out with different parameters, e.g., magnification. 

\begin{figure}
    \centering
    \includegraphics[width=1\linewidth]{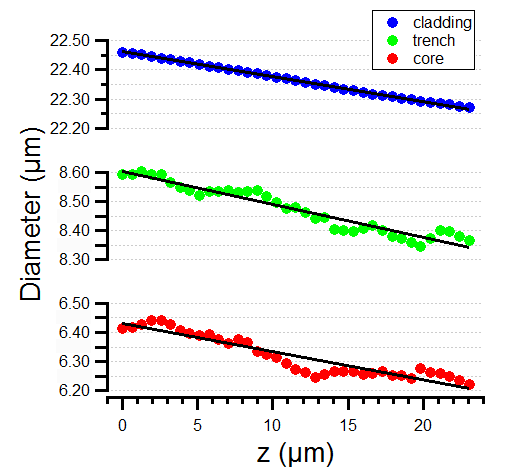}
    \caption{Evolution of the SMF taper cladding, trench, and core vs. z. The solid black lines are linear fit: $D_{clad} = 22.46 -0.01 z$, $D_{trench} = 8.60 -0.01 z$, $D_{core} = 6.43 -0.01 z$. $R^2_{clad}=0.99$, $R^2_{trench}=0.94$, $R^2_{core}=0.87$.}
    \label{fig:taper}
\end{figure}

Finally, the core, trench, and cladding diameters of the SMF taper varies along $z$ as shown in Fig. \ref{fig:taper}. 
In this regard, it is important to note that the tapering process is supposed to keep the same proportionality between the core and the cladding sizes. This was experimentally verified: the values of the core, trench, and cladding diameter vary along z with the same slope (cfr. the linear fit in Fig. \ref{fig:taper}).

\begin{table*}[!htb]
    \centering
    \begin{tabular}{ccccccc}
\toprule
     \multicolumn{1}{c}{} & \multicolumn{2}{c}{GRIN} & \multicolumn{2}{c}{STEP} & \multicolumn{2}{c}{SMF}\\
       \hline
        & $D_{clad}$ & $D_{core}$ & $D_{clad}$ & $D_{core}$ & $D_{clad}$ & $D_{core}$\\
        \hline
        $\mu$CT & 127.26 $\pm$ 4.54 & 54.54 $\pm$ 4.54 & 131.81 $\pm$ 4.54 &  59.09 $\pm$ 4.54 & 127.26 $\pm$  4.54 & // \\
       XRM & 124.99 $\pm$ 0.95 & 53.48 $\pm$ 0.95 & 124.99 $\pm$ 0.95 &  51.52 $\pm$ 0.95 & 124.99 $\pm$ 0.95 & 10.49 $\pm$ 0.95 \\
\bottomrule
    \end{tabular}
    \caption{Diameters evaluated starting from the profiles in Fig. \ref{fig:diam_prof}.}
    \label{tab:diam_prof}
\end{table*}

\begin{table*}[!htb]
    \centering
    \begin{tabular}{ccccccc}
\toprule
     \multicolumn{1}{c}{\textcolor{white}{nulla}} & \multicolumn{2}{c}{GRIN} & \multicolumn{2}{c}{STEP} & \multicolumn{2}{c}{SMF}\\
       \hline
       & $D_{clad}$ & $D_{core}$ & $D_{clad}$ & $D_{core}$ & $D_{clad}$ & $D_{core}$\\
        \hline
        $\mu$CT & 125.41 $\pm$ 1.79 & 49.02 $\pm$ 0.32 & 125.75 $\pm$ 1.37 & 51.83 $\pm$ 1.85 & 122.54 $\pm$ 0.58 & // \\
       XRM & 124.70 $\pm$ 0.16 & 49.84 $\pm$ 0.15 & 123.81 $\pm$ 0.18 & 49.87 $\pm$ 0.42 & 124.21 $\pm$ 0.24 & 8.44 $\pm$ 0.24 \\
\bottomrule
    \end{tabular}
    \caption{Diameters evaluated starting from segmentation.}
    \label{tab:diam_seg}
\end{table*}

We also computed $D_{core}$ and $D_{clad}$ starting from segmentation. Specifically, the threshold on the gray value borders a region, with an area $A$, that identifies either the core or the cladding. As such, the diameter was calculated as
\begin{equation}
    D = 2 \sqrt{A/\pi}.
\end{equation}
The results are reported in Table \ref{tab:diam_seg}. As can be seen, the results are in good agreement with those obtained starting from the $I_\mu$ profile.

As can be seen in Table \ref{tab:diam_prof} and \ref{tab:diam_seg}, $\mu$CT and XRM provide similar results. Indeed, the values of $D_{core}$ and $D_{clad}$ of all fibers are compatible, within the experimental error. However, the size of the SMF core, which turned out to be of the order of 10 $\mu$m with XRM measurements, was too small to be detected by $\mu$CT. As such, the value of its diameter is missing in Table \ref{tab:diam_prof} and \ref{tab:diam_seg}.

Setting a threshold on the XCT intensity for segmentation also allowed us to estimate the core and cladding non-circularity as
\begin{equation}
    \textit{NC} = \frac{S_{max}}{S_{min}}-1,
\end{equation}
where $S_{max}$ and $S_{min}$ are the maximum and the minimum sizes (say, diameter) of the segmented area. \textcolor{black}{Specifically, the different components (core, trench, and cladding) of the fiber were virtually separated. Then,  using Fiji software, we carried out a slice-by-slice fit using an elliptical shape, which provided the values of $S_{max}$ and $S_{min}$.}

\begin{figure}
\centering
\includegraphics[width=8.0cm]{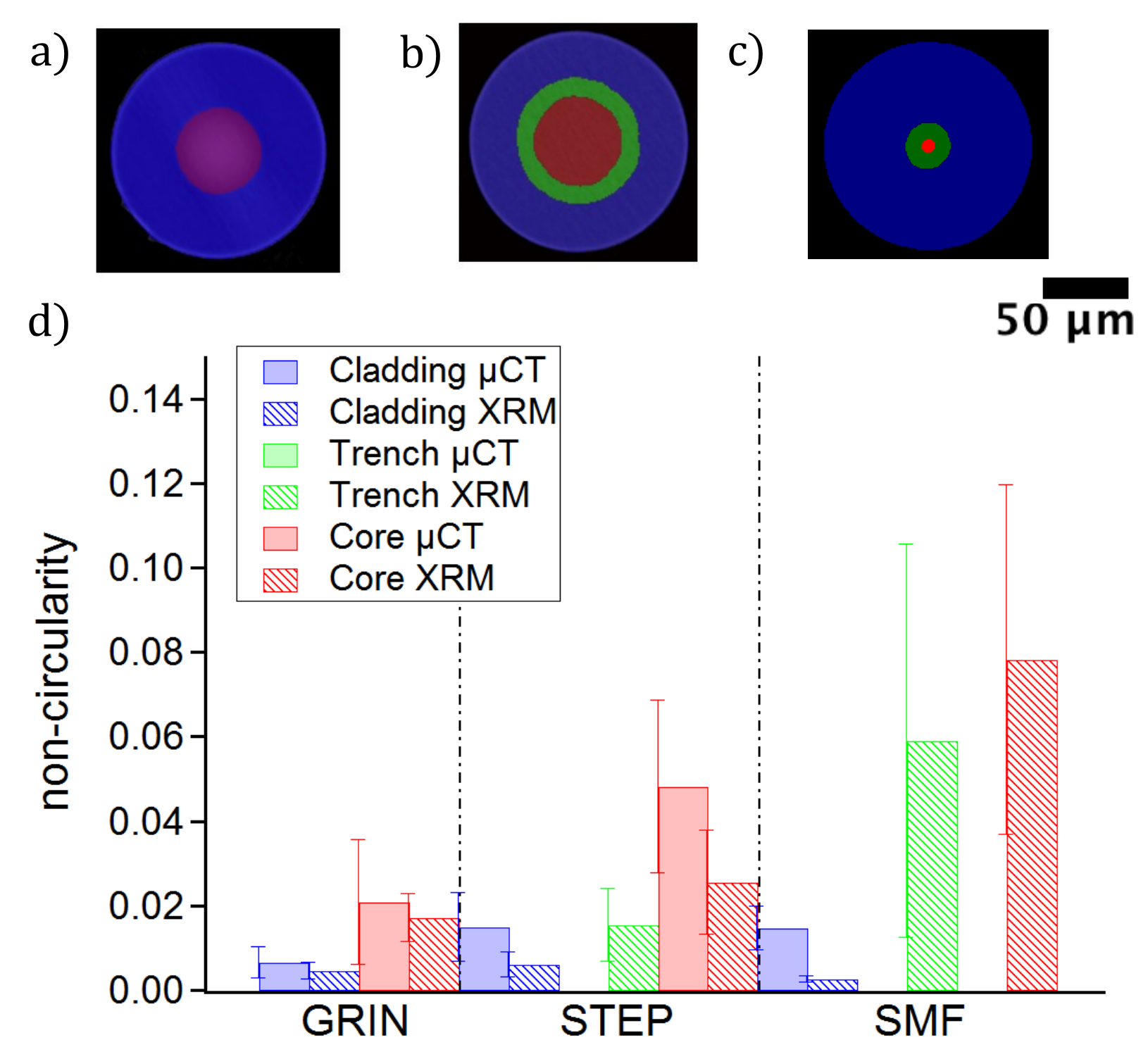}
\caption{Fiber non-circularity. (a-c) Segmentation of a single slice of the GRIN (a), STEP (b), and SMF (c) sample, respectively, obtained by XRM. The blue, green, and red colors refer to the cladding, trench, and core, respectively. (d) Histogram of the non circularity values.}
\label{fig:circ}%
\end{figure}

The results of segmentation and non-circularity estimation are shown in Fig. \ref{fig:circ}. Specifically, Fig. \ref{fig:circ}a-c show the segmentation of the XRM images of the GRIN, STEP, and SMF samples, respectively. Whereas, in Fig. \ref{fig:circ}d, we plot the values of the non-circularity that we found for each segmented area, i.e., core, cladding, and trench (except of the GRIN fiber), using both $\mu$CT and XRM. 

As can be seen in the latter, the values of \textit{NC}${}_{clad}$ are lower than those of \textit{NC}${}_{core}$ and \textit{NC}${}_{\textit{trench}}$ in all samples. This is consistent with the fact that the trench and the core are likely made via modified chemical vapor deposition. Thus, the core and trench are more sensitive to diameter fluctuations than the cladding, whose size can be finely tuned with mechanical processes. 

For the sake of readability, in Table \ref{tab:circularity} we report the maximum values of \textit{NC}, obtained as its average value plus its standard deviation calculated over 300 slices. As can be seen, the values that we found are in excellent agreement with those provided by the manufacturer (cfr. Table \ref{tab:samples}). In this regard, we emphasize that the geometrical parameters of the trench were not reported in Table \ref{tab:diam_prof}, \ref{tab:diam_seg} and \ref{tab:circularity} since we could not compare them with the nominal values provided by the manufacturer.

\begin{table*}[!ht]
    \centering
    \begin{tabular}{ccccccccc}
\toprule
     \multicolumn{1}{c}{\textcolor{white}{nulla}} & \multicolumn{2}{c}{GRIN} & \multicolumn{2}{c}{STEP} & \multicolumn{2}{c}{SMF} & \multicolumn{2}{c}{SMF Taper}\\
       \hline
       \textcolor{white}{nulla} & \textit{NC}${}_{clad}$ & \textit{NC}${}_{core}$ & \textit{NC}${}_{clad}$ & \textit{NC}${}_{core}$ & \textit{NC}${}_{clad}$ & \textit{NC}${}_{core}$ & \textit{NC}${}_{clad}$ & \textit{NC}${}_{core}$\\
        \hline
        $\mu$CT & $\leq 1.04\%$ & $\leq 3.57\%$ & $\leq 2.32\%$ & $\leq 6.87\%$ & $\leq 1.98\%$ & // & - & - \\
       XRM & $\leq 0.67\%$ & $\leq 2.27\%$ & $\leq 0.91\%$ & $\leq 3.79\%$ & $\leq 0.34\%$ & $\leq 11.98\%$ & - & - \\
       nCT & - & - & - & - & - & - & $\leq 0.16\%$& $\leq 2.86\%$ \\
\bottomrule
    \end{tabular}
    \caption{Fiber non-circularity evaluated starting from segmentation as in Fig. \ref{fig:circ}.}
    \label{tab:circularity}
\end{table*}

\section{Conclusion}
\label{sec:con}

In conclusion, by correlating X-ray absorption data with refractive index properties, XCT enables a comprehensive understanding of how fiber geometry and material composition influence optical behavior. Thanks to its non-destructive nature, XCT is particularly well-suited for analyzing optical fibers in various conditions, including coated, bent, or deformed states.  

In this work, we carried out a multiscale approach to XCT of standard optical fibers using classical sources. Our analysis highlights the strengths and limitations of different XCT methods, i.e., $\mu$CT, XRM, and nCT, specifically in terms of resolution, field of view, and segmentation efficiency. Furthermore, the integration of DL-assisted thresholding proves to be an effective tool for segmenting XCT images, mitigating the impact of artifacts and enhancing the accuracy of geometrical parameters evaluation. \textcolor{black}{We determined the diameter and the non-circularity of the fiber core and cladding, finding excellent agreement with those provided by the manufacturers. In particular, our estimation of the fiber diameter and non-circularity turned out to be affected by a typical relative uncertainty as low as 2\%.}
Each XCT technique analyzed in this study offers distinct advantages depending on the spatial scale of interest. In Tab. \ref{tab:advantages}, we summarize the target types of optical fibers associated with each XCT method.

\begin{table}[!hb]
    \centering
    \begin{tabular}{ll}
\toprule
        & Target fiber type \\
\midrule
       $\mu$CT  & Large core \\
        & long spans\\
\midrule
       XRM  & Small core\\
        & microstructured\\
\midrule
       nCT  & Nanofibers \\
\midrule
       Synch.  & Laser-induced damages\\
\bottomrule
    \end{tabular}
    \caption{Overview of target applications of $\mu$CT, XRM, nCT, and synchrotron XCT for optical fiber samples.}
    \label{tab:advantages}
\end{table}

In short, $\mu$CT is particularly useful for examining long fibers, even when wound on spools. Its wide field of view \textcolor{black}{(of the order of 10 x 10 mm\textsuperscript{2})} allows for the study of already drawn fibers, enabling the assessment of refractive index fluctuations along the fiber axis. However, this comes at the cost of reduced resolution\textcolor{black}{, which is of merely a few microns}. 

On the other hand, nCT provides the highest level of detail, making it ideal for investigating structures \textcolor{black}{as fine as 100 -200 nm}. However, the method turns out to be ineffective for standard 125 $\mu$m fibers, owing to its small field of view\textcolor{black}{, which is limited to 60 $\mu$m}.

Finally, XRM emerges as the most promising technique, as it combines high resolution \textcolor{black}{(about 1-2 $\mu$m)} with a sufficiently large field of view \textcolor{black}{(1 x 1 mm\textsuperscript{2})}. This makes XRM a strong candidate for the quality control of fibers even during the drawing process, as well as for fiber-based optical components such as optical fiber couplers. In this regard, AI-driven approaches can make a substantial impact by accelerating image acquisition, reconstruction, and segmentation, ultimately enhancing the efficiency and accuracy of fiber manufacturing. Indeed, AI-driven approaches could enable real-time monitoring and adaptive process control, optimizing fiber production while maintaining high-quality standards. 

Generally speaking, as fiber technologies continue to evolve, the adaptability of XCT, combined with AI-driven enhancements, will make it a cornerstone technique for both researchers and manufacturers in the field of optical fiber development. In particular, we foresee that XRM will be a powerful characterization tool for industrial fiber towers, whose long fiber span drawing is more likely to be affected by environmental perturbations.

In perspective, by leveraging phase contrast, we foresee that synchrotron XCT can be used to analyze small features inside optical fibers, e.g., submicron-scale defects created by lasers or splicing, with exceptional precision.

\subsubsection*{Acknowledgments}
This work was supported by the Italian Ministry of University and Research grant Progetto STAR 2, PIR01-00008 and by the "Advanced Tomography and Microscopies" (ATOM) Project, granted by Lazio Region (Prot. \#173-2017-17395 L.R. 13/2008), Regional call "Open Infrastructures for Research", by Piano Nazionale di Ripresa e Resilienza. (PNRR)—Research Infrastructure Project iENTRANCE\@ENL (www.ientrance.eu) "Infrastructure for Energy Transition and Circular Economy \@ EuroNanoLab" granted by Italian Ministry of University and Research (MUR), (Prot \#IR0000027, call 3264, Dec. 28 2021), and by the by European Union – Next Generation EU in the framework of the National Recovery and Resilience Plan through the research project I-PHOQS - Integrated Infrastructure Initiative in Photonic and Quantum Sciences (B53C22001750006).
We acknowledge the support of Sandro Donato and Fabio Mangini for sample preparation and image analysis.

\printcredits

\bibliographystyle{elsarticle-num-names}
\bibliography{cas-refs}

\end{document}